\documentclass{aastex631}

\begin{document}

\title{Two-dimensional Parameter Relationships for W UMa-type Systems Revisited}

\author{Atila Poro}
\altaffiliation{poroatila@gmail.com}
\affiliation{Astronomy Department of the Raderon AI Lab., BC., Burnaby, Canada}

\author{Ehsan Paki}
\affiliation{Astronomy Department of the Raderon AI Lab., BC., Burnaby, Canada}

\author{Ailar Alizadehsabegh}
\affiliation{Faculty of Electrical and Computer Engineering, Department of Photonics, University of Tabriz, Tabriz 51664, Iran}

\author{Mehdi Khodadadilori}
\affiliation{Department of Electrical Engineering, Amirkabir University of Technology, Tehran, Iran}

\author{Selda Ranjbar Salehian}
\affiliation{Astronomy Students’ Scientific Association, University of Tabriz, Tabriz 51664, Iran}

\author{Mahya Hedayatjoo}
\affiliation{Department of Physics, Iran University of Science and Technology, Tehran, Iran}

\author{Fatemeh Hashemi}
\affiliation{Institute for the Intellectual Development (Kanoon), Astronomy Group, Bojnord, Iran}

\author{Yasaman Dashti}
\affiliation{Department of Mechanical Engineering, Amirkabir University of Technology, Tehran, Iran}

\author{Fatemeh Mohammadizadeh}
\affiliation{Aerospace Engineering Department, Science and Research Branch of Islamic Azad University, Tehran, Iran}

\begin{abstract}
Reviewing the empirical and theoretical parameter relationships between various parameters is a good way to understand more about contact binary systems. In this investigation, two-dimensional (2D) relationships for $P-M_{V(system)}$, $P-L_{1,2}$, $M_{1,2}-L_{1,2}$, and $q-L_{ratio}$ were revisited. The sample used is related to 118 contact binary systems with an orbital period shorter than 0.6 days whose absolute parameters were estimated based on the $Gaia$ Data Release 3 (DR3) parallax. We reviewed previous studies on 2D relationships and updated six parameter relationships. Therefore, Markov chain Monte Carlo (MCMC) and Machine Learning (ML) methods were used, and the outcomes were compared. We selected 22 contact binary systems from eight previous studies for comparison, which had light curve solutions using spectroscopic data. The results show that the systems are in good agreement with the results of this study.
\end{abstract}

\keywords{Close binary stars – Fundamental parameters of stars – Astronomy data analysis}

\vspace{1cm}
\section{Introduction}
The W Ursae Majoris (W UMa) stars are one of the most interesting types of binary systems, and they are important astrophysical tools for understanding star formation, structure, and evolution. Both stars in these contact binary systems have exceeded their Roche lobes, and mass transfer through Lagrange points is likely to occur.
The W UMa-type systems are known as Low-Temperature Contact Binaries (LTCBs), and the difference between the temperatures of two components in these types of eclipsing stars is close to equal (\citealt{2005ApJ...629.1055Y}).

W UMa binary systems typically have orbital periods of less than one day, but different intervals are specifically defined for contact binaries (e.g., \citealt{2003MNRAS.342.1260Q}, \citealt{2017RAA....17...87Q}, \citealt{2017RAA....17..115J}, \citealt{2018PASJ...70...90K}, \citealt{2018ApJ...859..140C}, \citealt{2020MNRAS.493.4045J}, \citealt{2021ApJS..254...10L}, \citealt{2022PASP..134f4201P}, \citealt{2022MNRAS.510.5315P}). However, according to the \cite{2021ApJS..254...10L} study, systems with an orbital period and temperature of more than 0.5 days and 7000 K are likely to have radiative envelopes and should not be classified as W UMa-type binaries.

Although there have been many theoretical and observational research publications, our knowledge of contact binary stars is still far from complete. The relationships between the absolute parameters in the W UMa-type systems continue to be ambiguous, especially in view of the various techniques for the analysis of light curves and the number of contact binary systems that have been studied. There are some strong and weak trends in the relationships between parameters; however, it appears that these results depend on the sample and the method used to study the relationships. For this reason, investigations of these relationships by using modern methods are always useful.
\\
\\
In this study, we continued to use a sample of absolute parameters for 118 contact systems estimated in the \cite{2022MNRAS.510.5315P} study. The aim is to investigate the relationship between some parameters in two-dimensional space. This paper’s structure is as follows: The data used to analyze the relationships between parameters is described in Section 2; the relationships between the orbital period, mass, and luminosity of the contact binary systems are discussed in Sections 3 and 4; updated relations and related methods are presented in Section 5; and finally, there is a conclusion in Section 6.

\vspace{1cm}
\section{Dataset}
The \cite{2022MNRAS.510.5315P} investigation employed 118 contact systems with orbital periods less than 0.6 days, and we used this sample and results in our study. These systems vary from 7 to 17 apparent magnitudes in the $V$ filter, and they are both A and W types of contact binaries from both hemispheres. Considering that both sub-types of W UMa are used in the sample, identifying which star is primary and which one is secondary is based on the study that is used as the reference study for each system. Reference papers include both types of photometric and spectroscopic studies.

In the \cite{2022MNRAS.510.5315P} study, the orbital period and the parameters obtained from the light curve solutions are taken from the literature, and absolute parameters for them are determined using the $Gaia$ DR3 parallax. The parameters $d$(pc), $A_V$, $V_{max}$(mag), $l_{1,2}/l_{tot}$, $BC_{1,2}$, $T_{1,2}$, $r_{mean1,2}$, and $P$(day) are needed for estimation of absolute parameters. $M_{V(system)}$, $M_{V1,2}$, $M_{bol1,2}$, $L_{1,2}$, $R_{1,2}$, $a_{1,2}$, $M_{1,2}$ calculated, respectively.

It is remarkable that \cite{2022MNRAS.510.5315P} have mentioned current research on the systematic zero-point offset of the $Gaia$ DR3 parallaxes and its values that need to be adjusted.

Based on the \cite{2022MNRAS.510.5315P} study, the obtained total mass compared to the total mass from the literature that used the photometric method shows about an 18 percent improvement over spectroscopic investigations. 
\\
\\
We obtained the absolute magnitude parameter of each system since it was required in our analysis. The absolute magnitude of each system is calculated using equation \ref{eq1}:

\begin{equation}\label{eq1}
M_{V(system)}=V-5log(d)+5-A_V
\end{equation}

where $V$ is the maximum apparent magnitude of the system, $d$(pc) is the distance of the system calculated by using $Gaia$ DR3\footnote{\url{https://gea.esac.esa.int/archive/}} parallax, the extinction coefficient is $A_v$ and its uncertainty was estimated using the DUST-MAPS PYTHON package of \cite{2019ApJ...887...93G}. The computation results are outlined in Table \ref{tab1}.

\begin{table*}
\caption{Calculated systems' absolute magnitudes of the sample.}
\centering
\begin{center}
\footnotesize
\begin{tabular}{c c c c c c c c}
 \hline
 \hline
 System & $M_V$ & System & $M_V$ & System & $M_V$ & System & $M_V$\\
\hline
ASAS J212234-4627.6	&	4.506(38)	&	AB And	&	5.608(58)	&	BH Cas	&	4.200(160)	&	MU Cnc	&	4.946(246)	\\
ASAS J212319-4622.4	&	4.896(169)	&	GZ And	&	4.354(67)	&	V573 Peg	&	3.664(214)	&	V596 Peg	&	5.216(221)	\\
ASAS J174406+2446.8	&	4.193(33)	&	AO Cam	&	4.779(382)	&	EP And	&	3.409(317)	&	V658 Lyr	&	3.942(58)	\\
1SWASP J064501.21+342154.9	&	6.626(342)	&	DK Cyg	&	2.559(236)	&	AP Leo	&	3.428(34)	&	V700 Cyg	&	4.920(102)	\\
2MASS 02272637+1156494	&	7.393(172)	&	KW Psc	&	6.242(151)	&	BI CVn	&	3.668(189)	&	FZ Ori	&	3.430(258)	\\
ASAS J083241+2332.4	&	3.896(52)	&	LO And	&	3.805(202)	&	AL Cas	&	2.995(52)	&	LP UMa	&	4.298(77)	\\
1SWASP J155822.10-025604.8	&	5.418(27)	&	V1191 Cyg	&	4.200(340)	&	V680 Per	&	3.399(106)	&	FP Lyn	&	3.882(126)	\\
1SWASP J212808.86+151622.0	&	6.954(29)	&	V1853 Ori	&	3.775(193)	&	EZ Hydrae	&	3.770(248)	&	FV CVn	&	4.584(36)	\\
UCAC4 436-062932	&	4.842(36)	&	CE Leo	&	5.558(22)	&	MQ UMa	&	2.513(64)	&	V354 UMa	&	4.418(74)	\\
TYC 1597-2327-1	&	5.359(197)	&	V532 Mon	&	3.119(181)	&	V1197 Her	&	6.163(50)	&	OQ Cam	&	3.220(257)	\\
1SWASP J080150.03+471433.8	&	6.398(378)	&	GU Ori	&	3.547(267)	&	AH Cnc	&	3.759(229)	&	EQ Tau	&	4.522(263)	\\
J015829.5+260333	&	3.500(127)	&	FI Boo	&	4.034(28)	&	V781 Tau	&	3.735(42)	&	V737 Per	&	4.622(295)	\\
J030505.1+293443	&	6.179(195)	&	AV Pup	&	3.21(64)	&	V1848 Ori	&	5.702(124)	&	V336 TrA	&	5.626(95)	\\
SDSS J012119.10-001949.9	&	7.443(71)	&	NN Vir	&	2.229(23)	&	QQ Boo	&	4.801(219)	&	HN Psc	&	4.158(86)	\\
1SWASP J140533.33+114639.1	&	6.453(212)	&	AQ Psc	&	2.893(80)	&	V608 Cas	&	3.961(232)	&	V685 Peg	&	4.795(305)	\\
ROTSE1 J164341.65+251748.1	&	4.581(381)	&	XY Leo	&	5.481(29)	&	QX And	&	3.005(116)	&	BO Ari	&	4.287(30)	\\
GSC 03526-01995	&	5.644(374)	&	V2284 Cyg	&	4.88(167)	&	UY UMa	&	3.539(163)	&	V351 Peg	&	1.849(38)	\\
GSC 3581-1856	&	5.493(166)	&	GK Aqr	&	5.100(248)	&	BF Pav	&	4.999(137)	&	AK Her	&	3.176(24)	\\
GSC 1042-2191	&	2.760(27)	&	V396 Mon	&	4.052(350)	&	AA UMa	&	3.489(148)	&	HI Dra	&	1.755(22)	\\
NSVS 2701634	&	5.968(152)	&	TT Cet	&	2.670(71)	&	NR Cam	&	5.385(167)	&	V1128 Tau	&	4.556(33)	\\
NSVS 2643686	&	3.398(338)	&	PS Vir	&	4.731(35)	&	V776 Cas	&	2.258(26)	&	V2612 Oph	&	3.789(39)	\\
NSVS 7245866	&	3.374(179)	&	BQ Ari	&	5.114(56)	&	PY Vir	&	5.287(83)	&	EP Cep	&	5.419(195)	\\
NSVS 1557555	&	5.127(50)	&	V870 Ara	&	3.546(79)	&	RZ Tau	&	3.307(105)	&	EQ Cep	&	5.507(170)	\\
GSC 2723-2376	&	5.080(68)	&	V811 Cep	&	5.550(121)	&	EH CVn	&	5.829(62)	&	ES Cep	&	4.107(153)	\\
GSC 4946-0765	&	5.276(43)	&	V842 Cep	&	4.674(311)	&	EF CVn	&	4.858(119)	&	V369 Cep	&	4.922(184)	\\
V2240 Cyg	&	2.734(82)	&	TZ Boo	&	4.239(171)	&	V584 Cam	&	3.497(121)	&	V370 Cep	&	4.635(160)	\\
V1370 Tau	&	4.568(177)	&	GW Boo	&	2.502(122)	&	GR Vir	&	3.929(22)	&	V782 Cep	&	4.342(74)	\\
V1007 Cas	&	4.631(190)	&	BE Cep	&	4.014(218)	&	FG Hydrae	&	4.130(104)	&	GW Cnc	&	5.105(245)	\\
V789 Her	&	4.677(250)	&	VZ Psc	&	6.413(339)	&	UX Eri	&	3.834(172)	&	DF CVn	&	4.820(256)	\\
NX Cam	&	2.141(96)	&	RT LMi	&	3.614(269)	&		&		&		&		\\
\hline
\hline
\end{tabular}
\end{center}
\label{tab1}
\end{table*}

\vspace{1cm}
\section{Period-Luminosity Relationship}
There are interesting features of W UMa-type systems that make them special for study. For instance, the existence of a common envelope to transfer mass and energy, the dependence of the radii ratio on the mass ratio through Kuiper’s relation (\citealt{1941ApJ....93..133K}), and sharing the same Roche surface. Apart from this, it is possible to derive an orbital period-luminosity or an orbital period-absolute magnitude (generally all called $P-L$) relationship for each contact system as a consequence of the common envelope. Since \cite{1967MmRAS..70..111E}’s investigations using $P-L$ relations contact binaries seemed to be useful as distance indicators, similar to those of the classical standard candles such as Cepheid stars. \cite{1994PASP..106..462R}, \cite{2004NewAR..48..703R}, and \cite{2016ApJ...832..138C} have published the framework for these relations. However, several investigations have reported a variety of results for contact binary systems and provided the period-luminosity relations.

\cite{1994PASP..106..462R} proposed an orbital period-luminosity-color relationship using 18 W UMa-type contact binaries, which was improved by the \cite{1997PASP..109.1340R} study by utilizing 40 W UMa-type and $HIPPARCOS$ parallaxes. \cite{2006MNRAS.368.1319R} constructed a $P-L$ relation using 21 nearby late-type contact binaries, although only 11 members of the sample followed it; therefore, the relation has a large uncertainty. \cite{2014MNRAS.443..432M} and \cite{2016MNRAS.457.4323P} obtained relations using a sample of eclipsing binaries in the Large Magellanic Cloud (LMC) which led to an unclear $P-L$ relation due to not correcting for the dominant period-color relations. In another study by \cite{2016ApJ...832..138C} the first reliable $10\%$ (1-sigma) accuracy, the $P-L$ relation was presented using 66 contact binaries that confirmed the use of W UMa-type binaries as distance indicators.

\cite{2017AJ....154..125M} investigated a $P-L$ relationship using Tycho-$Gaia$ astrometric solution (TGAS) parallaxes. They found a steep linear relationship between the absolute magnitude and the orbital period in the orbital period range of 0.22 to 0.88 days. In order to improve upon the previous studies \cite{2018ApJ...859..140C} used 183 nearby W UMa-type contact binaries with accurate TGAS parallaxes for present $P-L$ relations. In the \cite{2018ApJ...859..140C} study photometric observations in the $G$ band and mid-infrared bands were also taken into consideration.

In a recent work \cite{2021ApJS..254...10L} have determined the dependence of the absolute parameters on the orbital period and mass ratio. In addition, relations that indicate the dependence are presented. \cite{2021ApJS..254...10L} used a sample of 210 W UMa-type systems with orbital periods $<0.5$ days and $L_1<5L_\odot$ for an investigation of luminosity against orbital period. A linear fit through data has determined the $P-L$ relations between the primary and secondary component stars. Table \ref{tab2} lists some of the relationships found by various investigators.

\begin{table*}
\caption{Some of the orbital period, mass, and luminosity relations presented by previous investigations for contact systems.}
\centering
\begin{center}
\footnotesize
\begin{tabular}{c c c}
 \hline
 \hline
 Parameters	& Relation	& References\\
\hline
$P-M_v$ & $M_v=(-12.0\pm2.0)logP+(-1.5\pm0.8)$ &\cite{2006MNRAS.368.1319R}\\
$P-M_V(max)$ & $M_V(max)=(-9.15\pm0.12)logP+(-0.23\pm0.05),$ & \cite{chen2016contact}\\
& $\sigma_V=0.30(logP<-0.25)$ &\\
$P-M_V(max)$ & $M_V(max)=(-4.95\pm0.13)logP+(0.85\pm0.02),$ & \cite{chen2016contact}\\
& $\sigma_V=0.35(logP>-0.25)$ &\\
$P-L_1$ & $L_1=(13.98\pm0.75)P-(3.04\pm0.27)$ &\cite{2021ApJS..254...10L}\\
$P-L_2$ & $L_2=(3.66\pm0.26)P-(0.69\pm0.09)$ &\cite{2021ApJS..254...10L}\\
$M_{ratio}-L_{ratio}$&	$L_2/L_1=(M_2/M_1)^{0.92}$	&\cite{1968ApJ...153..877L}\\
$M_{ratio}-L_{ratio}$&	$L_2/L_1=(M_2/M_1)^{0.96}$	&\cite{1968ApJ...151.1123L}\\
$M_{ratio}-L_{ratio}$&	$L_2/L_1=(M_2/M_1)^{0.92}$	&\cite{1970AA.....5...12C}\\
$M_{ratio}-L_{ratio}$&	$L_2/L_1=(M_2/M_1)^{0.93}$	&\cite{1981AAS...46..193H}\\
$M_{ratio}-L_{ratio}$&	$L_2/L_1=(M_2/M_1)^{0.82}$	&\cite{1992ApSS.189..237R}\\
$M_{ratio}-L_{ratio}$&	$L_2/L_1=(M_2/M_1)^{1.04}$	&\cite{1992ApSS.189..237R}\\
$M_{ratio}-L_{ratio}$&	$L_2/L_1=(M_2/M_1)^{4.60}$	& \cite{2004AA...426.1001C}\\
$M_{ratio}-L_{ratio}$&	$L_2/L_1=(M_2/M_1)^{0.74}$	&\cite{2005JKAS...38...43A}\\
$M_1-L_1$ &	$logL_1=(2.92\pm0.11)logM_1+(0.01\pm0.02)$ & \cite{2021ApJS..254...10L}\\
$M_2-L_2$ &	$logL_2=(0.69\pm0.09)logM_2+(0.13\pm0.05)$ & \cite{2021ApJS..254...10L}\\
\hline
$P-M_V-Color$ & $M_V=-4.44logP+3.02(B-V)_0+0.12$ & \cite{rucinski1997absolute}\\
$P-M_V-Color$ & $M_V=-4.43logP+3.63(V-I)_0-0.31$ & \cite{rucinski2004contact}\\
$P-M_V-Color$ & $M_V(mean)=-5.20logP+1.19(G-W1)_0-0.09
$ & \cite{chen2018optical}\\
\hline
\hline
\end{tabular}
\end{center}
\label{tab2}
\end{table*}

\vspace{1cm}
\section{Mass-Luminosity Relationship}
W UMa stars are contact binaries with a common convective envelope in which the photosphere stands between the inner and outer Lagrangian zero-velocity surfaces (\citealt{1968ApJ...153..877L}, \citealt{1968ApJ...151.1123L}). The primary star in contact binary systems transfers its mass and luminosity to the secondary one throughout the common envelope between the stars. The light curve shape and the character of the mass ratio-luminosity ratio relation of W UMa-type stars are explained by assuming the first accepted contact model by \cite{1968ApJ...153..877L} and \cite{1968ApJ...151.1123L}. In the mentioned model the luminosity transfers from the more massive star to the less massive one while the two stars are in thermal and geometrical equality. The model explains that both stars touch each other in spite of filling their Roche-lobe. \cite{1981ApJ...245..650M} assumed that the energy transfer rate is solely determined by the mass ratio, whereas \cite{2001OAP....14...33K} proposed that the energy transfer rate in W UMa-type systems is determined by the luminosity of the secondary star. The following assumptions by \cite{2002AA...395..899K} and \cite{2002AA...395..907K} show that the rate of the transferred luminosity is variable in time.

Further investigation by \cite{2004AA...426.1001C} proved that this transfer parameter is a function of both mass and luminosity ratios. They also proved that different types of contact binaries are located in distinct areas on the mass ratio-luminosity ratio diagram. In addition, they found that with respect to energy transfer, systems with mass ratios higher than $q=0.72$ form an individual group on the diagram which is introduced as the H-subtype contact binary. Furthermore, a model of W UMa binaries was suggested by \cite{2009MNRAS.397..857S} in which the exchanged energy in the components is attained through large-scale circulation in the equatorial plane. This is found to be in agreement with the recent observational results of \cite{2015AJ....149...49R} and \cite{2020AJ....160..104R} on the structural and evolutionary theories of contact binaries. On the other hand, these theories try to explain the very small differences in the effective temperatures of the components of W UMa binary systems. \cite{2020MNRAS.492.4112Z} also tried to explain the W UMa phenomenon in terms of stellar evolution in contact systems.

A mass-luminary relationship in contact binaries has been considered the most substantial characteristic behavior of the W UMa-type stars due to the fact that it leads to an understanding of stellar structure and evolution in these systems. The relation was defined simply as $L\propto M^a$ in which $L$ represents the luminosity ratio and $M$ is the mass ratio. \cite{1965PASJ...17...97O} showed that the mass ratio and the luminosity ratio in those systems should be equal ($L_2/L_1=M_2/M_1$) with $a=1$. \cite{1968ApJ...153..877L} found that the relation between the observable luminosity ratio and the ratio of the stellar surfaces is $L_2/L_1=(M_2/M_1)^{0.92}$. According to the literature, this relation has been improved by several studies (Table \ref{tab2}).

\vspace{1cm}
\section{Updated Relations and Methods}
We employed both ML and MCMC methods to obtain the best fit and compare the results.

a) This study used the ML regression approach to find the best polynomial fit for each parameter set. Polynomial regression is a form of regression analysis (\citealt{shaeri2022prediction}). The relationship between the independent variable $x$ and the dependent variable $y$ is not linear but is the polynomial degree of nth in $x$ (\citealt{2011JMLR...12.2825P}). The equation for polynomial regression is also derived from the linear regression equation (polynomial regression of degree 1), which is defined as:

\begin{equation}
\label{eq2}
y=b_0+b_1 x+b_2 x^2+...+b_n x^n+e
\end{equation}

where $y$ is the predicted/target output, $b_0+b_1,...b_n$ are the regression coefficients, and $x$ is an independent input variable. We can say that if data are not distributed linearly but instead represent the nth degree of a polynomial, then we use polynomial regression to get the desired output.

A polynomial regression model consists of successive power terms, and degree is the most important parameter in polynomial regression. We use the Bootstrap technique to find the best degree parameter for each relation. The bootstrap is a versatile technique that relies on data-driven simulations to make statistical inferences. When combined with robust estimators the bootstrap can afford much more powerful and flexible inferences than is possible with standard approaches such as t-tests on means. We calculated the Root Mean Square Error (RMSE) for each order and identified the best fit to find the best order for the fractional model. One thousand rounds are used for each relation with 30\% of the data as a testing set. The minimum average of the RMSE of all 1000 running rounds in each degree parameter was used as a basis to select the best degree parameter for the polynomial regression of each relation. RMSE can be expressed as:

\begin{equation}
\label{eq3}
RMSE=\sqrt{\sum_{i=1}^{n}{\frac{(\hat{y}_i-y_i)^2}{n}}}
\end{equation}

where $n$ is the number of data points, $y_i$ is the $i-th$ measurement, and $\hat{y}_i$ is its corresponding prediction.
Best-fit values of the equation parameters are determined from this paper's dataset using a polynomial regression and parameter error calculated by Standard Error Mean (SEM). The dataset was split by 15\% for training and validation. The regression model is implemented in Python and uses Polynomial Features data preprocessing. The linear regression model is available in the Python Scikit-learn library (\citealt{2011JMLR...12.2825P}).

b) We also used the MCMC approach to fit polynomial models to the data (\citealt{2018ApJS..236...11H}). We applied MCMC to estimate the posterior probability function and the distribution of parameters, which depend on the polynomial coefficients and are consistent with the dataset. The optimization was done by the Python library SciPy v1.10.0. We sampled the distribution using 50 walkers and ran the MCMC for 20000 steps using the emcee Python package (\citealt{2013PASP..125..306F}). The MCMC uncertainties are derived from the $16th$, $50th$, and $84th$ percentiles of the samples in the marginalized distributions.

\begin{figure*}
\begin{center}
\includegraphics[scale=0.35]{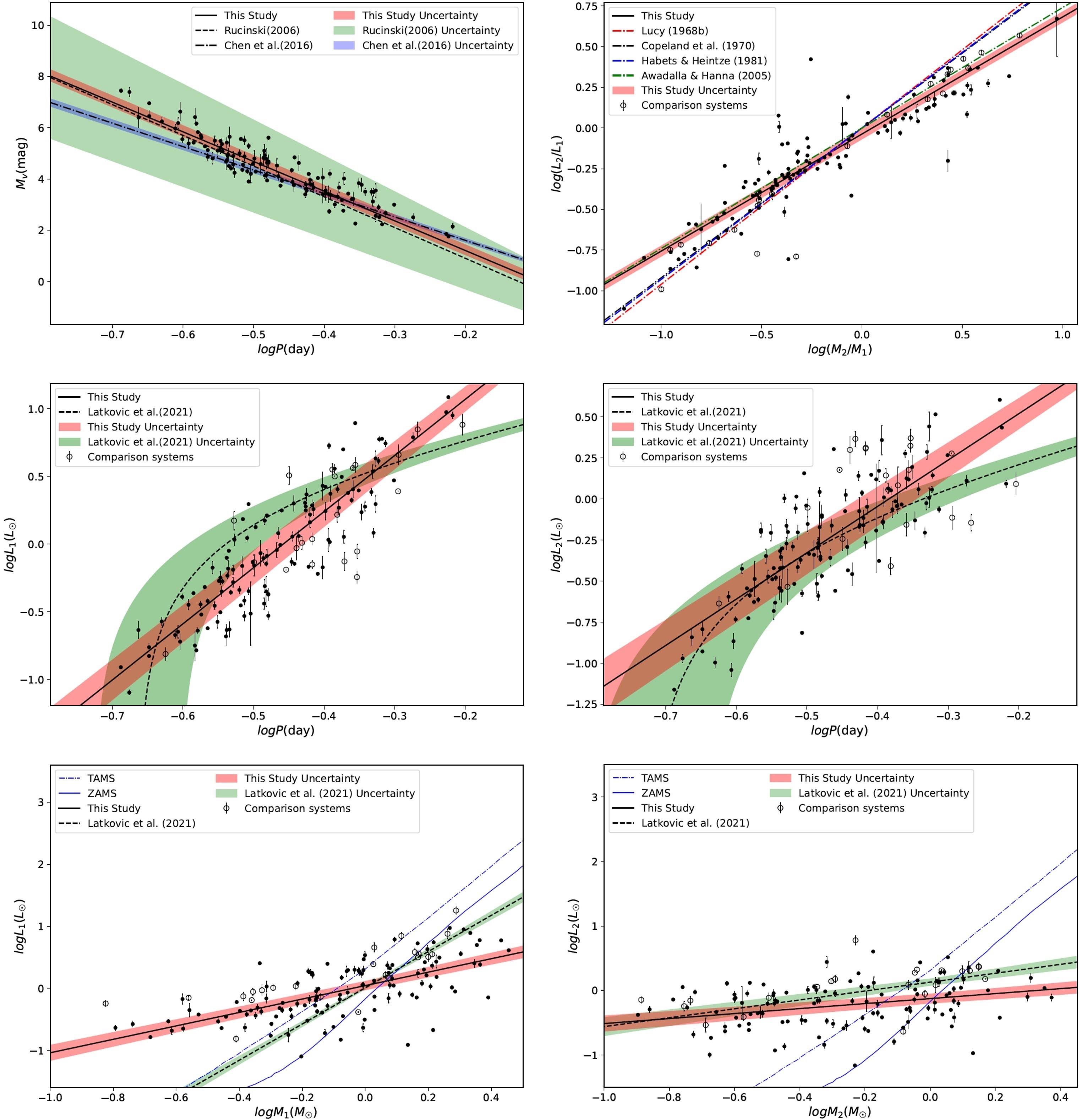}
    \caption{2D relationships between different parameters in the W UMa-type binary systems. The filled circles correspond to the sample used in this study (118 systems), and the hollow circles are the 22 contact binary systems for comparison. The relations $M_{1,2}-L_{1,2}$ are shown together with the theoretical zero-age main sequence (ZAMS) and terminal-age main sequence (TAMS).}
\label{Fig1}
\end{center}
\end{figure*}

\begin{figure*}
\begin{center}
\includegraphics[scale=0.32]{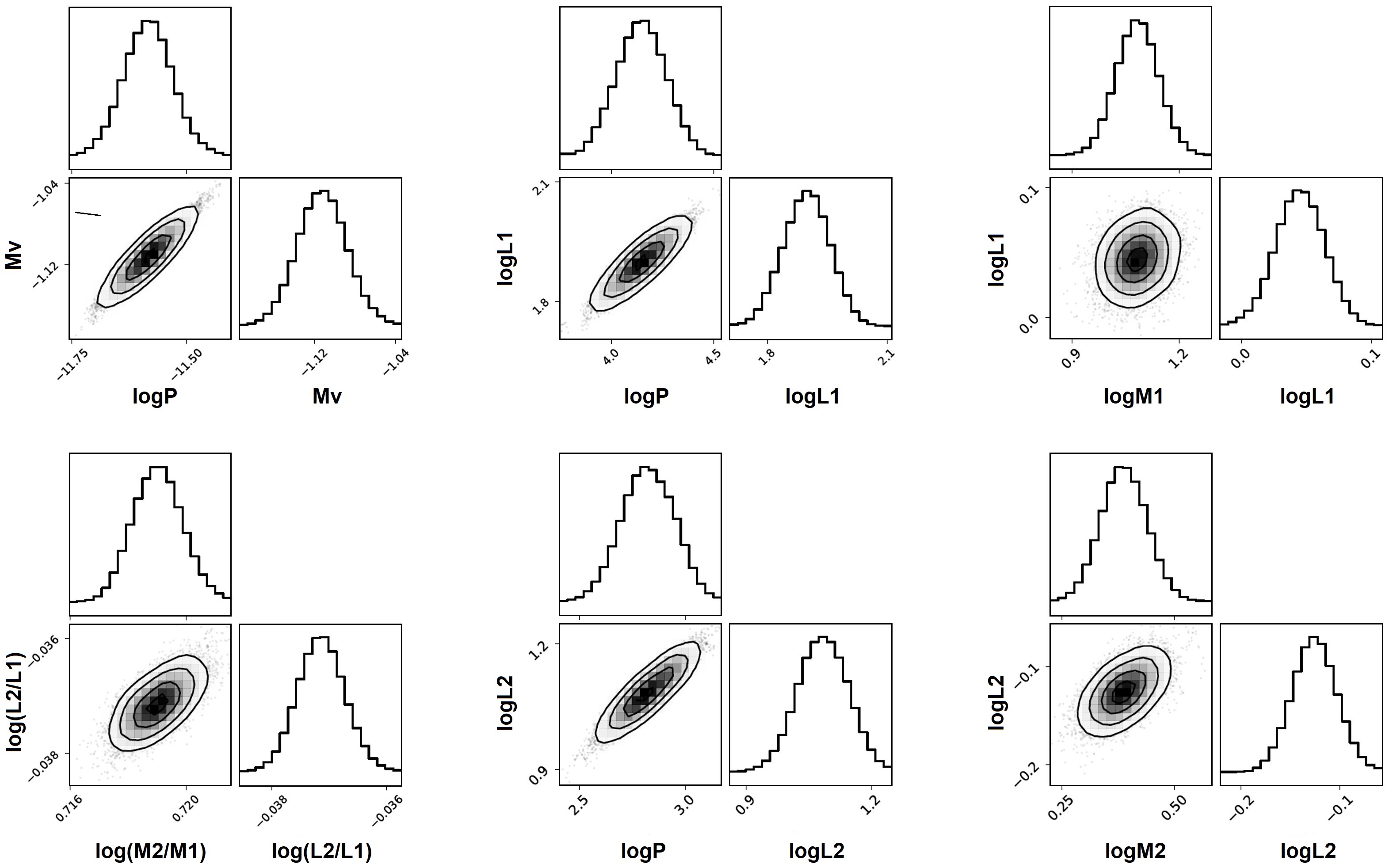}
    \caption{Corner plots of the posterior distribution based on the MCMC sampling.}
\label{Fig2}
\end{center}
\end{figure*}

2D relationships are displayed in Figure \ref{Fig1} between $P-M_V$, $P-L_{1,2}$, $M_{1,2}-L_{1,2}$, and $q-L_{ratio}$. Figure \ref{Fig2} shows the MCMC corner plots of 1D and 2D projections of the posterior probability distributions of polynomial coefficients, and we can see coefficient dependency.
\\
\\
The results of the two MCMC and ML methods were compared, and it was found that the relationships were precisely the same, with an average difference in uncertainty of only one percent. As a result, we decided to present new relationships using the MCMC approach (Equations \ref{eq4} to \ref{eq9}). However, RMSE for each relationship based on the ML method are as follows: $P-L_1=0.187$, $P-L_2=0.168$, $M_1-L_1=0.323$, $M_2-L_2=0.302$, $P-M_{V(system)}=0.125$, and $q-L_{ratio}=0.113$.

Some 2D relations (Table \ref{tab2}) presented by the previous studies were shown in Figure \ref{Fig1}.
We displayed the $P-M_V$ fit related to the \cite{2006MNRAS.368.1319R} study in Figure \ref{Fig1} to illustrate that a significant amount of uncertainty has been taken into account for it.

Also, in the relationship of $q-L_{ratio}$ in Figure 1, we also displayed four relations from previous studies for comparison. The relationship between $q-L_{ratio}$ obtained from this study can also be used as $\cong(L_2/L_1)=(M_2/M_1)^{0.72}$ to be comparable with other relationships in Table \ref{tab2}.

In Figure \ref{Fig1}, five 2D relationships also showed the results of the \cite{2021ApJS..254...10L}  study. According to the objectives of the \cite{2021ApJS..254...10L} study, the more massive star is set as star 1, which is different from our sample.
Anyway, the logarithmic scale of the relationship between $L_1-M_1$ presented in the \cite{2021ApJS..254...10L} study reveals that there is a considerable difference compared to this study. However, the 22 comparison contact systems (Figure \ref{Fig1}) for $L_1-M_1$ provide a better agreement with the fit of our investigation. Generally, the samples showed more scatter in the two relations $L_1-M_1$ and $L_2-M_2$ than in other relationships.

It should be noted that the two relationships $P-L_{1,2}$ in the \cite{2021ApJS..254...10L} study were not presented on a logarithmic scale. These two relationships do not show a linear fit when converted to logarithmic form. According to \cite{2021ApJS..254...10L} study's sample, the logarithmic scale could be a more appropriate choice for the orbital period interval that they considered.

\begin{equation}
\label{eq4}
M_V=(-11.58\pm 0.05)\times logP+(-1.11\pm 0.66)
\end{equation}

\begin{equation}
\label{eq5}
logL_1=(4.15\pm 0.11)\times logP+(1.90\pm 0.13)
\end{equation}

\begin{equation}\label{eq6}
logL_2=(2.82\pm 0.12)\times logP+(1.08\pm 0.15)
\end{equation}

\begin{equation}
\label{eq7}
logL_1=(1.08\pm 0.06)\times
logM_1+(0.05\pm 0.04)
\end{equation}

\begin{equation}\label{eq8}
logL_2=(0.39\pm 0.05)\times logM_2+(-0.13\pm 0.06)
\end{equation}

\begin{equation}\label{eq9}
log\frac{L_2}{L_1}=(0.72\pm 0.01)\times log\frac{M_2}{M_1}+(-0.04\pm 0.02)
\end{equation}

\vspace{1cm}
\section{Discussion and Conclusion}
1. Investigating the relationships between different parameters can be helpful for improving our understanding of W UMa-type systems. As it is clear in Table \ref{tab2}, in previous studies, investigations have been done with different samples and methods and attempts have been made to reveal the relationships between parameters. This study revisits the relationships between some parameters in two-dimensional space related to contact binary stars using modern methods.

2. A sample of 118 contact binary systems from the \cite{2022MNRAS.510.5315P} study was selected. The selection of a sample is important for updating empirical and theoretical relationships. In our sample, the $Gaia$ DR3 parallax method was used to estimate the absolute parameters from the light curve solutions of the selected studies. It should be noted that in the $Gaia$ DR3 parallax method for contact systems if some parameters from the light curve solution have a problem, the values of $a_1(R_{\odot})$ and $a_2(R_{\odot})$ will not be close to each other and they cannot be averaged to obtain $a(R_{\odot})$.

There are some other samples that have been used in previous investigations on contact binary systems' parameter relationships. These samples were gathered to be utilized in those studies, and some changes have been made, such as normalizing the mass ratio and considering more massive stars as primary, which cannot be compared with the current investigation. Some systems in these samples have written the wrong values compared to their study. However, the most important issue to consider is the lack of uniformity in the methods used to estimate absolute parameters in different studies that are considered for samples.

3. We focused on the period-luminosity and mass-luminosity relationships. First, the results of previous studies were reviewed, and then according to our sample and methods, the best fits were determined. MCMC and ML methods were used for this purpose, and comparing their results showed that they were the same. Consequently, updated parameter relationships were presented for $P-M_{V(system)}$, $P-L_{1,2}$, $M_{1,2}-L_{1,2}$, and $q-L_{ratio}$ (Equations \ref{eq4}-\ref{eq9}). Figure \ref{Fig1} suggests that among these six 2D relationships, $P-L_{1,2}$, $P-M_{V(system)}$, and $q-L_{ratio}$ have stronger relationships whereas $M_{1,2}-L_{1,2}$ has the weakest.

4. We used a logarithmic scale for the six relationships. Algorithmic scales allow for greater flexibility in analyzing the relationship between parameters. According to performing and comparing relationships with two methods of machine learning and MCMC in this study, the logarithmic scale increases accuracy. Shorter intervals in the logarithmic scale provide the possibility of a larger number of searches for these two methods, and the best fit could be obtained. Although at first we also checked all relationships with non-logarithmic scales, some relationships in this case required polynomial relationships, but with a logarithmic scale, there was a possibility of a linear fit.

5. Based on the estimation of absolute parameters, $M-L$ is one of the diagrams that shows the evolutionary stage of each star in a contact binary system.
The primary star will be near the ZAMS considering the components in W UMa-type eclipsing binaries share a Common Connective Envelope (CCE) (\citealt{2005ApJ...629.1055Y}). Usually, primary stars should be more around ZAMS, and we have seen secondary stars near TAMS and above it (Figure \ref{Fig1}). According to the \cite{2013MNRAS.430.2029Y} study, the trend is completely opposite for these two stars in W-type systems, which explains why some stars can be located in unexpected locations in $M_{1,2}-L_{1,2}$ diagrams (Figure \ref{Fig1}).

6. We selected different 22 W UMa-type systems than our sample with periods shorter than 0.6 days from eight studies for comparison (Table \ref{tab3}). These systems have analysis with spectroscopic data and also from both A and W types. The stars of the considered systems are in good agreement with the fits and our sample data points, as shown in Figure \ref{Fig1}. Considering that the absolute magnitude parameter of the systems is not estimated in the studies, it was not possible to display and compare it with the $P-M_{V(system)}$ relationship of this study.

\begin{table*}
\caption{Selected 22 contact binary systems were analyzed with spectroscopic data to evaluate parameter relationships.}
\centering
\begin{center}
\footnotesize
\begin{tabular}{c c c c c c c c c}
 \hline
 \hline
 System	& Type & $P(day)$	& $q=M_2/M_1$	& $M_1(M_{\odot})$	& $M_2(M_{\odot})$	& $L_1(L_{\odot})$	& $L_2(L_{\odot})$ &	Reference
\\
\hline
QX And	& A &	0.412172	&	0.306(9)	&	1.470(50)	&	0.450(20)	&	3.160	&	1.140	&	\cite{2011AA...525A..66D}	\\
RW Com	& W &	0.237346	&	2.123(6)	&	0.389(20)	&	0.826(54)	&	0.154(17)	&	0.231(23)	&	\cite{2011MNRAS.412.1787D}	\\
OU Ser	& A &	0.296768	&	0.173(17)	&	1.187(99)	&	0.205(20)	&	1.488(245)	&	0.292(81)	&	\cite{2011MNRAS.412.1787D}	\\
V2357 Oph	& A &	0.415568	&	0.231(10)	&	1.160(61)	&	0.268(25)	&	1.646(220)	&	0.390(51)	&	\cite{2011MNRAS.412.1787D}		\\
CK Boo	& A &	0.355152	&	0.111(52)	&	1.584(34)	&	0.176(8)	&	3.211(518)	&	0.572(101)	&	\cite{2011MNRAS.412.1787D}		\\
XX Sex	& A &	0.540108	&	0.100(2)	&	1.301(22)	&	0.130(6)	&	7.021(909)	&	0.717(88)	&	\cite{2011MNRAS.412.1787D}		\\
TW Cet	& W &	0.316851	&	1.334(30)	&	0.722(53)	&	0.963(97)	&	0.733(97)	&	0.886(115)	&	\cite{2011MNRAS.412.1787D}		\\
V2377 Oph	& W &	0.425406	&	2.532(12)	&	0.410(71)	&	1.038(192)	&	0.744(120)	&	1.212(355)	&	\cite{2011MNRAS.412.1787D}		\\
BV Eri	& A &	0.507655	&	0.300(20)	&	1.068(123)	&	0.320(57)	&	4.561(822)	&	0.770(135)	&	\cite{2011MNRAS.412.1787D}		\\
VY Sex	& W &	0.443433	&	3.195(5)	&	0.440(19)	&	1.406(80)	&	0.882(119)	&	2.345(308)	&	\cite{2011MNRAS.412.1787D}		\\
LS Del	& W &	0.363842	&	2.666(10)	&	0.470(138)	&	1.253(401)	&	0.933(206)	&	1.992(403)	&	\cite{2011MNRAS.412.1787D}		\\
V417 Aql	& W &	0.370314	&	2.765(7)	&	0.510(31)	&	1.410(113)	&	1.020(117)	&	2.327(260)	&	\cite{2011MNRAS.412.1787D}		\\
TX Cnc	& W &	0.382883	&	2.196(11)	&	0.602(20)	&	1.322(57)	&	1.088(112)	&	2.030(189)	&	\cite{2011MNRAS.412.1787D}		\\
SZ Hor	& A &	0.625102	&	0.470(40)	&	1.823(159)	&	0.857(147)	&	7.605(1.419)	&	1.234(203)	&	\cite{2011MNRAS.412.1787D}		\\
V839 Oph	& W &	0.409008	&	0.305(24)	&	1.630(24)	&	0.497(16)	&	3.550(350)	&	1.392(138)	&	\cite{2011MNRAS.412.1787D}		\\
DY Cet	& A &	0.440790	&	0.356(9)	&	1.436(34)	&	0.511(25)	&	3.840(482)	&	1.507(194)	&	\cite{2011MNRAS.412.1787D}		\\
V535 Ara	& A &	0.629306	&	0.302(3)	&	1.940(40)	&	0.590(20)	&	18(3)	&	6(1)	&	\cite{2012NewA...17..143O}	\\
V546 And	& W &	0.383020	&	3.937(11)	&	0.275(8)	&	1.083(30)	&	0.706(45)	&	2.052(41)	&	\cite{2015NewA...36..100G}	\\
KIC 10618253	& A &	0.437403	&	0.125(1)	&	1.476(22)	&	0.184(3)	&	3.639(63)	&	0.698(116)	&	\cite{2016PASA...33...43S}	\\
OO Aql	& A &	0.506792	&	0.846	&	1.060(7)	&	0.897(6)	&	2.453(7)	&	1.894(6)	&	\cite{2016RAA....16....2L}	\\
V972 Her	& W &	0.443094	&	6.099(88)	&	0.150(10)	&	0.910(70)	&	0.570(60)	&	2.110(190)	&	\cite{2018ApSS.363...34S}	\\
NSVS 4161544	& W &	0.351680	&	3.377(5)	&	0.436(4)	&	1.473(7)	&	0.646(4)	&	1.506(9)	&	\cite{2019AJ....157...73K}	\\
\hline
\hline
\end{tabular}
\end{center}
\label{tab3}
\end{table*}

\vspace{1cm}
\section*{Acknowledgements}
The Binary Systems of South and North (BSN) project (\url{https://bsnp.info/}) and Raderon AI Lab's Astronomy Department provided scientific support for this study (\url{https://astronomy.raderonlab.ca/}).
We have made use of data from the European Space Agency (ESA) mission $Gaia$ (\url{http://www.cosmos.esa.int/gaia}), processed by the $Gaia$ Data Processing and Analysis Consortium (DPAC). We would like to thank Golnaz Mazhari, Edwin Budding, and Paul D. Maley for their scientific assistance.

\vspace{1cm}
\section*{ORCID iDs}
\noindent Atila Poro: 0000-0002-0196-9732\\
Ehsan Paki: 0000-0001-9746-2284\\
Ailar Alizadehsabegh: 0000-0001-5768-0340\\
Mehdi Khodadadilori: 0000-0002-9967-6144\\
Selda Ranjbar Salehian: 0000-0002-5223-1332\\
Mahya Hedayatjoo: 0000-0002-0192-215X\\
Fatemeh Hashemi: 0000-0002-0248-1202\\
Yasaman Dashti: 0000-0001-8442-6305\\
Fatemeh Mohammadizadeh: 0000-0002-9552-6667\\

\vspace{1cm}

\bibliography{sample631}{}
\bibliographystyle{aasjournal}

\end{document}